\documentclass{article}
\usepackage{graphicx}
\title{VRXU-net: A Deep Learning Approach for Brain Ischemic Stroke Lesion Detection and Segmentation in T1W MRI}
\author{Sayed Amir Mousavi Mobarakeh}
\date{May 2026}
\usepackage{booktabs}
\usepackage{amsmath,amssymb,amsfonts}
\usepackage{multirow}

\usepackage{algorithmic}
\usepackage{graphicx}
\usepackage{textcomp}
\usepackage{graphicx}
\usepackage[flushleft]{threeparttable}
\begin{document}

\maketitle

\begin{abstract}
When the blood supply to the brain is obstructed by a clot, oxygen delivery to brain tissues becomes insufficient, leading to cellular necrosis. In healthcare settings, accurately identifying and delineating ischemic lesion boundaries is essential for treatment and surgical planning. However, ischemic stroke lesions vary widely in shape, size, and location, and in grayscale MRI modalities such as T1W they may resemble surrounding brain structures. This makes lesion detection and segmentation a challenging task for clinicians.
This study introduces a novel VRU-Net architecture, derived from visual features, residual connections, and a U-shaped network, for detecting and segmenting ischemic stroke lesions in 3D magnetic resonance imaging scans. The proposed method first uses a modified VGG model to identify ischemic stroke in separate 2D slices. Then, a U-shaped segmentation model with residual blocks segments the lesion in each slice. This procedure is applied independently to the axial, sagittal, and coronal planes, and the final output is generated by aggregating the three segmentation results.
To improve both performance and processing speed, a high-performance classifier is applied before the segmentation model in a sequential framework. This strategy reduces unnecessary segmentation of non-lesion slices and improves overall accuracy. In addition, decomposing 3D images into 2D slices reduces model complexity while allowing information from three anatomical planes to support more accurate lesion localization.
The proposed model is trained on the Anatomical Tracings of Lesions After Stroke dataset and outperforms state-of-the-art models in terms of accuracy and Dice coefficient. Moreover, the segmentation output provides feedback that helps the classification model reduce false-positive predictions.

\end{abstract}
Ischemic Stroke, T1W, Classification, Segmentation, Machine learning
\section{Introduction}

\label{sec:introduction}
Stroke, also referred to as a cerebrovascular accident, is a prevalent neurological disorder worldwide. Additionally, stroke ranks as the second primary contributor to mortality. The severity of the condition can manifest in varying degrees of dysfunction, disability, and mortality, thus emphasizing the gravity of the disease \cite{1}, \cite{2}. According to the 2022 Global Stroke Factsheet \cite{3}, there has been a noteworthy upsurge of 50\% in the lifetime risk of experiencing a stroke during the past 17 years. Presently, it is approximated that one-quarter of the general population will undergo a stroke in their lifetime.

Thus, cerebral stroke has a significant socioeconomic impact on society and a momentous effect on the quality of life of each individual patient \cite{4}. Among the most frequently observed symptoms in individuals affected by stroke are dysphasia, dysarthria, hemianopia, weakness, and seizures \cite{5}.
Cerebral Stroke can be categorized into two distinct types, namely ischemic stroke and hemorrhagic stroke. The former involves the obstruction of blood flow caused by the formation of blood clots, leading to a consequential deprivation of oxygen supply to brain cells, culminating in cellular death. Conversely, the latter results from the rupture of blood vessels within the brain, leading to bleeding and an associated rise in pressure on brain tissue \cite{3}. Globally, ischemic stroke is the most prevalent form of stroke, comprising approximately 80\% of all reported stroke cases \cite{4,5}. The classification of ischemic stroke is contingent upon the duration of time elapsed since symptom onset, with acute, subacute, and chronic phases representing the initial day, one to two weeks, and more than two weeks post-onset, respectively\cite{6}. During the acute stage of ischemic stroke, the resulting lesion can be classified into three distinct regions based on their potential for recovery, collectively referred to as "salvageability": the core, penumbra, and benign oligemia \cite{7}. The core region, situated at the lesion's center, denotes irreparable damage. The penumbra region, located around the core, is hyperperfused, indicating that the brain tissue in this area is at risk but still viable for recovery. The benign oligemia is a slightly hyperperfused outer ring, with no risk to the brain tissue in this area \cite{8}. \\

ATLAS

\section{Related work}
Classification and segmentation techniques have been widely employed in various fields and contexts, with a particular emphasis on biomedical imaging, including ischemic stroke. This section presents a concise summary of the progression of image classification and segmentation within the domain of medical imaging, specifically in regards to its application in diagnosing and analyzing ischemic stroke.\\
Saatman et al. \cite{19} have explored various techniques for the classification of traumatic brain injuries, with Glasgow Coma Scale (GCS) serving as the key determinant of injury severity. Meanwhile, in another study \cite{20}, the authors proposed an approach for identifying abnormalities in CT images, categorizing them into three distinct classes: chronic, hemorrhage, and acute infarct.
The method's limitations include a lack of reported accuracy and comparison with additional advanced approaches. Additionally, Griffis et al. \cite{21} have employed Gaussian naive Bayes classification as a predictive tool to identify the region impacted by stroke in T1-weighted MRI scans, albeit with the inability to detect small white matter (WM) lesions. Shahangian and Pourghassemi \cite{22} have presented a method for classifying and segmenting hemorrhages, utilizing two approaches: modifying distance regularized level set evolution (MDRLSE) and support vector machine (SVM) classifiers for segmentation and classification, respectively. On the other hand, Subudhi et al. \cite{23} have employed expectation maximization and random forest classifiers in their study. Besides the aforementioned classification approaches, conventional machine learning techniques have also shown promising results. For instance,In the year 2012, Krizhevsky and colleagues proposed a neural network architecture known as AlexNet \cite{24}. AlexNet consisted of five convolutional layers and three fully connected layers, and it demonstrated exceptional performance by achieving first place in the ImageNet Large Scale Visual Recognition Challenge (ILSVRC). Subsequently, in 2014, Simonyan and Zisserman proposed a deeper architecture, named VGG-16 \cite{25}, that delivered notable performance on large datasets. An additional noteworthy structure is ResNet50 \cite{26}, developed by He et al., and emerged victorious in the ILSVRC competition of 2015. Several studies have proposed classical methods for ischemic segmentation. For example, in citation \cite{27}, the authors presented a technique based on K-Nearest Neighbor (KNN) that utilized voxel intensity and spatial information to segment white matter. In \cite{28}, a method was developed that utilized the multimodality of MRI images, and segmented the ischemic lesion by considering the symmetry of the brain. 
O Maier \cite{6} introduced a model called different tree forests and reported a dice score of 0.65\%. 
Chyzhyk \cite{30} developed a classifier model for voxel-based lesion segmentation that addressed the problem of different images contracting in multi-sequential images. In \cite{31} the author proposed a method comprising three steps: (i) the deployment of heuristic histogram equalization to improve image resolution; (ii) extraction of texture from images; and (iii) utilization of an adaptive neuro-fuzzy inference system (ANFIS) to classify images. McKinley \cite{32} introduced the modified random forest method and segmented medical images. Lastly, Maier \cite{33} suggested an approach that combined random forest and voxel-based segmentation.\\
The literature discussed in this study achieved significant results; however, recent advancements in convolutional neural networks (CNN) have substantially impacted image classification and segmentation and CNN has undergone rapid development in recent years, leading to the proposal of various methods and architectures. One of the most renowned architectures is U-net \cite{34}, which was proposed for image segmentation, especially for biomedical images, and yielded conventional results. For instance, Clerigue \cite{35} proposed a model based on U-net to segment acute and subacute ischemic lesion stroke, which was trained on multimodality MRI.
In one study \cite{37}, the author proposed a method that combined two models, one of which was trained on 2D images and the other on 3D images. The model was called dimension-fusion-Unet (D-Unet), and it achieved a dice score of 53\%. Another study proposed a method called Multi-Scale Deep Fusion Network (MSDF-Net) with Atrous Spatial Pyramid Pooling (ASPP) to extract features from images of chronic segment stroke, which resulted in a dice score of 56\% \cite{38}. Additionally, another proposed model in a study \cite{39} is based on a convolutional neural network named Res-CNN, which was trained on multimodality MRI and achieved an 88.43\% dice score. Another study proposed the deep residual attention convolution neural network (DRA net) method to segment and quantify ischemic stroke lesions \cite{40}. Liang proposed an approach in \cite{4} that comprised of two primary components. The first one involves an ensemble of two DeconvNets that are used to detect the lesion in images. The second one is called the multi-scale convolutional label evaluation net (MUS-CLE Net) and is responsible for evaluating the performance of the first part. The approach presented in \cite{41} involves first segmenting the tumor and then classifying it into one of four classes, and is based on a CNN network with cascaded deep learning. The model was trained on 2D images. Meanwhile, Konstantinos \cite{42} proposes a novel architecture that comprises two paths: one extracts features from a small context, while the other extracts features from a larger context. Qi et al. \cite{43} proposed a novel approach called X-Net, which incorporates a Feature Similarity Module (FSM) as a nonlocal operation to capture long-range dependencies. The FSM significantly improves the model's performance by effectively extracting dense contextual information. Consequently, the model achieves a Dice coefficient of 0.487\%. Yang et al. \cite{44} proposed the Cross-Level fusion and Context Inference Network (CLCI-Net) to segment chronic stroke lesions from T1-weighted MR images. The CLCI-Net incorporates a Cross-Level feature Fusion (CLF) strategy to effectively integrate features at different scales and levels. By extending the Atrous Spatial Pyramid Pooling (ASPP) with CLF, the method enhances the representation of multi-scale features, enabling the handling of lesions of varying sizes. Additionally, Convolutional Long Short-Term Memory (ConvLSTM) is employed to capture contextual information and address the challenge of intensity similarity, thereby improving the delineation of fine structures. The proposed method achieves a Dice score of 0.581\%.
The different loss functions are utilized in image segmentation. Binary-cross entropy is the famous loss function for image segmentation in binary mode \cite{45}. 
In several image segmentation tasks, the Dice coefficient is utilized as a loss function \cite{46}.

\begin{figure}
\centerline{\includegraphics[width=\columnwidth]{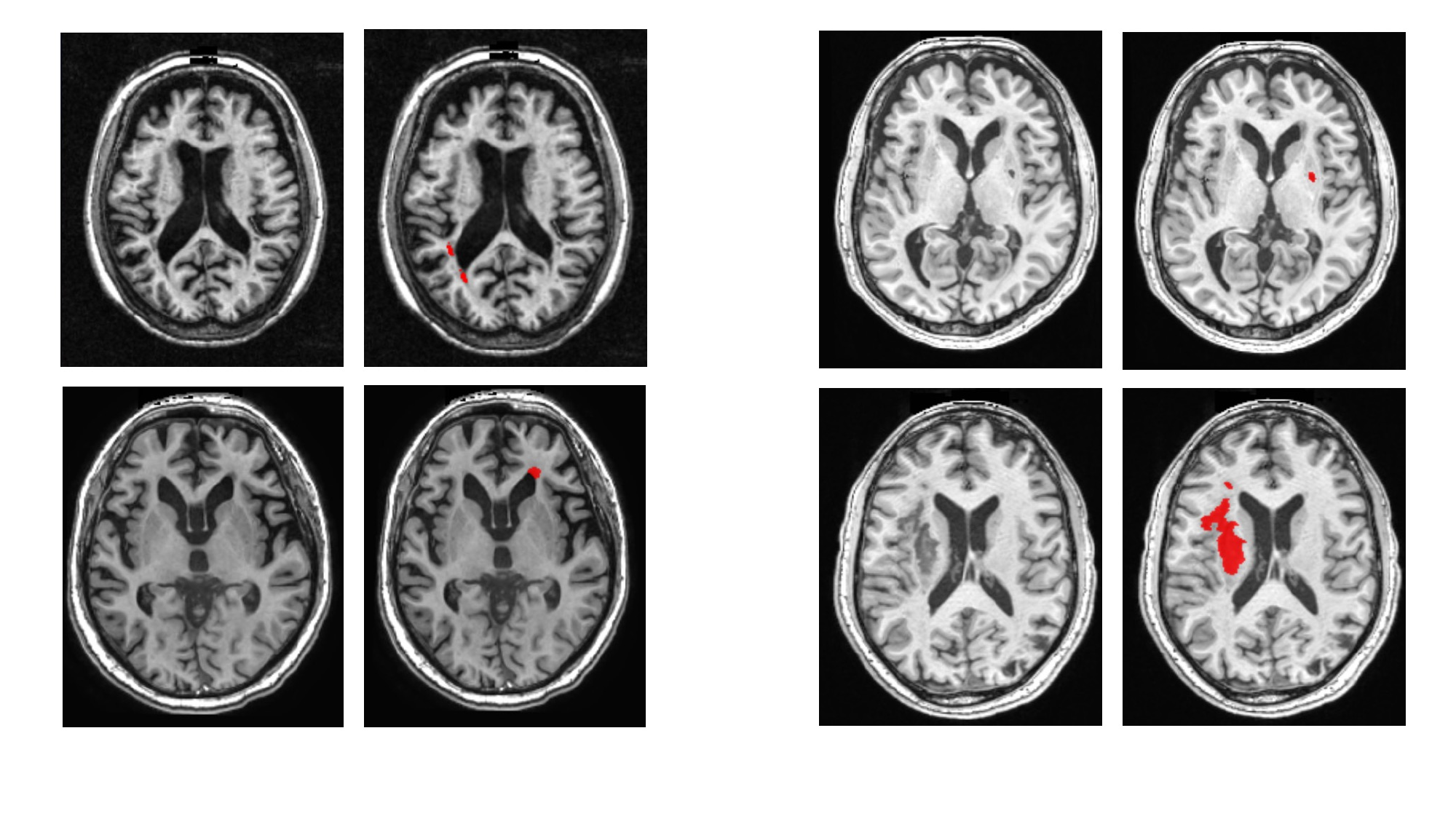}}
\caption{The stroke image derived from the ATLAS dataset corresponds to the MRI T1 sequence.The first and third columns represent the raw data. The second and fourth columns depict the raw data alongside the corresponding manually marked lesion boundaries, as annotated by the physician.}
\label{fig1}
\end{figure}

\section{data material}
In machine learning, data is the foundation upon which models are built and predictions are made. The performance of the model is directly influenced by the caliber and volume of data employed during the training process. The publicly accessible Anatomical Tracings of Lesions After Stroke (ATLAS) dataset serves as a repository of cerebral images and comprehensive lesion data pertaining to individuals who have experienced strokes. Developed by a collaborative group of researchers affiliated with the University of California, San Francisco, this dataset was conceived to furnish a uniform collection of lesion information that can be leveraged for diverse research purposes.
\subsection{Dataset}
The ATLAS dataset, designed for research purposes, offers a comprehensive compilation of ischemic stroke MRI scans, accompanied by accurately annotated lesion masks. This dataset encompasses a total of 955 cases, with 655 subjects having their corresponding ground truth labels available. Stringent quality control measures were applied to each MRI scan, ensuring reliability. Notably, each of the 655 patients possesses a single lesion mask file in three-dimensional (3D) format, which underwent thorough scrutiny by two teams of trained experts. It is important to mention that all subjects share identical sizes, shapes, and are represented in a single modality, specifically, T1-weighted. The data acquisition process involved the utilization of both 1.5 Tesla and 3-Tesla MRI scanners. Furthermore, the dataset features high-resolution data, typically with a minimum voxel size of 1 mm³ or greater. However, there exist four distinct groups of data that possess a resolution ranging from 1-2 mm³ for at least one dimension.

\begin{table}[h]
\centering
\caption{LESION NUMBER AND LOCATION IN HEMISPHERE}
\label{table}
\setlength{\tabcolsep}{0.1pt}
\begin{tabular}{ccccccc}
\toprule
\multicolumn{1}{c}{} & \multicolumn{3}{c}{\textbf{One Lesion}} & \multicolumn{3}{c}{\textbf{Multiple Lesions}} \\
\cmidrule(rl){2-4} \cmidrule(rl){5-7}
\textbf{} & {Left} & {Right} & {Other} & { Unilateral} & { Bilateral} & {Other}\\
\midrule
\centering $\begin{array}{c}  \textbf {Data} \\
\end{array}$ & 
\centering $\begin{array}{c}\\173 \\
(26.4 \%)\\ \text{} \end{array}$ & $\begin{array}{c}\\187 \\
(28.5 \%)\\ \text{}\end{array}$ & $\begin{array}{c}\\46 \\
(7.0 \%)\\ \text{}\end{array}$ & $\begin{array}{c}\\47 \\
(7.2 \%)\\ \text{}\end{array}$ & $\begin{array}{c}\\121 \\
(18.5 \%)\\ \text{}\end{array}$ & $\begin{array}{c}\\81 \\
(12.4 \%)\\ \text{}\end{array}$ \\

\bottomrule
\end{tabular}
\label{tab3}
\end{table}

\begin{table}[h]
\centering
\caption{LESION NUMBER AND LOCATION IN CORTICAL AND SUBCORTICAL}
\label{table}
\setlength{\tabcolsep}{0.1pt}
\begin{tabular}{ccccccc}
\toprule
\multicolumn{1}{c}{} & \multicolumn{2}{c}{\textbf{One Lesion}} & \multicolumn{2}{c}{\textbf{Multiple Lesions}} & \multicolumn{1}{c}{\textbf{Others}} & \multicolumn{1}{c}{\textbf{Total}} \\
\cmidrule(rl){2-3} \cmidrule(rl){4-5} 
\textbf{} & {Left} & {Right} & {Left} & {Right} & {} & {}\\
\midrule

$\begin{array}{l}\text {Data } \\
\end{array}$ & $\begin{array}{c}\\132 \\
(12.0 \%)\\ \text{}\end{array}$ & $\begin{array}{c}\\149 \\
(13.5 \%)\\ \text{}\end{array}$ & $\begin{array}{c}\\333 \\
(30.2 \%)\\ \text{}\end{array}$ & $\begin{array}{c}\\324 \\
(29.4 \%)\\ \text{}\end{array}$ & $\begin{array}{c}\\163 \\
(14.8 \%)\\ \text{}\end{array}$ & $\begin{array}{c}\\1101 \\
\\ \text{}\end{array}$ \\

\bottomrule
\end{tabular}
\label{tab3}
\end{table}

\begin{figure}[!t]
\centerline{\includegraphics[width=\columnwidth]{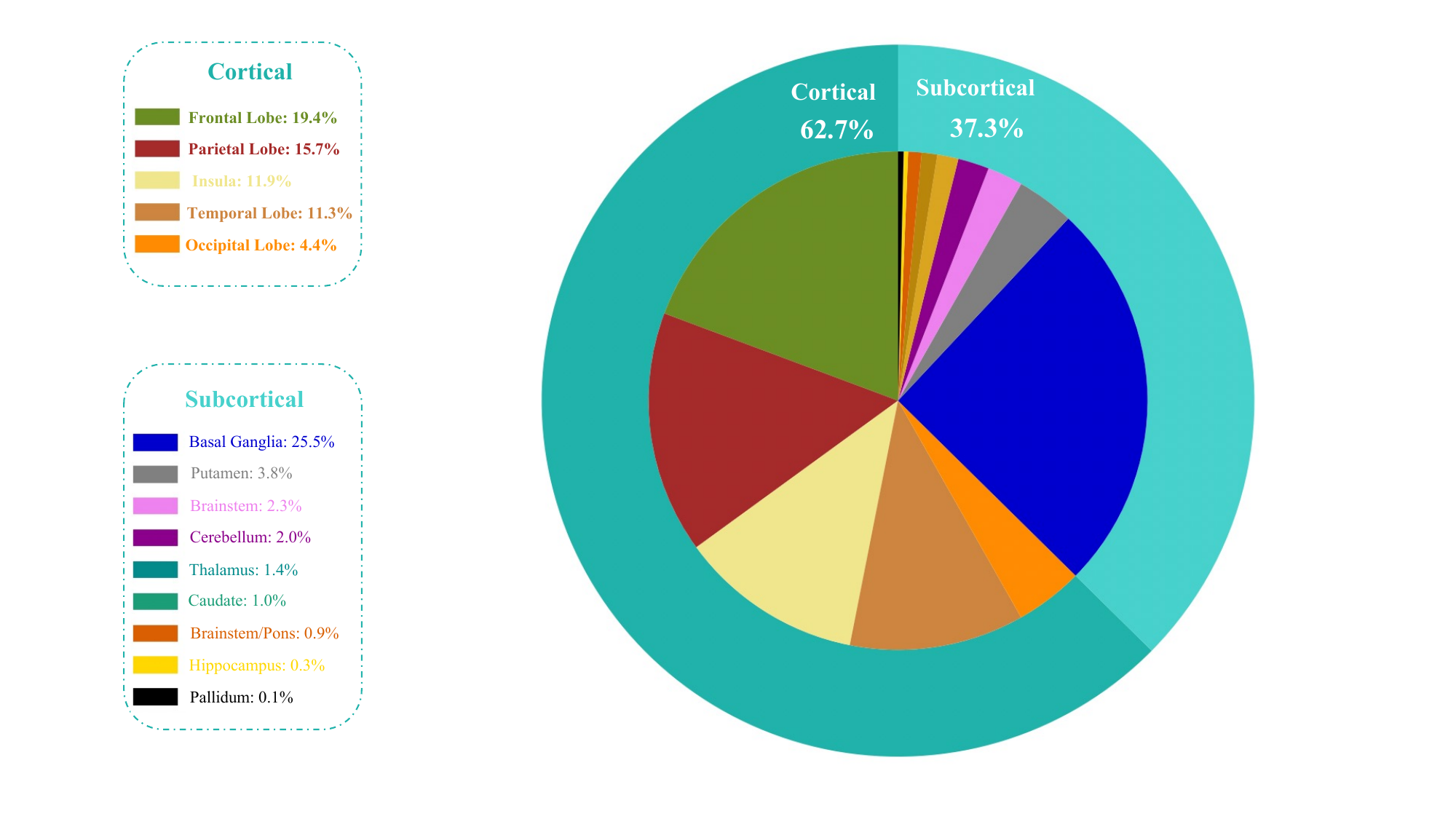}}
\caption{Lesion Distribution in Cortical and Subcortical Areas}
\label{fig1}
\end{figure}

\subsection{Lesion Distribution Patterns}
Considering the variability observed in brain anatomy, it is reasonable to posit that the lesions are distributed across diverse regions, and certain subjects may exhibit multiple lesions. During the evaluation stage of each lesion mask, a proficient team member systematically documented metadata, encompassing the number of lesions present and their respective localization in terms of lateralization (left or right hemisphere) and categorization as cortical or subcortical.
\subsubsection{Hemisphere Location}
Out of a total of 655 participants, the majority of 61.9\% exhibit a single lesion, while the remaining 38.1\% have multiple lesions. For subjects with only one lesion, 26.4\%, 28.5\%, and 7.0\% of them have the stroke lesion situated in the left hemisphere, right hemisphere, or other areas of the brain, respectively. On the other hand, for those with multiple lesions, the lesions can be either unilateral (7.2\%), bilateral (18.5\%), or in other locations (12.4\%), such as the cerebellum or brainstem. The terms "bilateral" and "unilateral" refer to the presence of lesions in either both hemispheres or only one hemisphere, respectively.
These findings are summarized in Table 1.

\subsubsection{Subcortical vs. Cortical Location}
Lesion distribution was categorized based on distinct anatomical regions within the brain, namely cortical, subcortical, and other areas, constituting 25.5\%, 59.6\%, and 14.8\% of the total lesions, respectively. Within the cortical region, 12\% and 13.5\% of the lesions were localized in the left and right hemispheres, respectively. In the subcortical region, 30.2\% and 29.4\% of the lesions were observed in the left and right hemispheres, respectively. The entire dataset encompassed a comprehensive collection of 1101 lesions. These insightful findings are concisely presented in Table 2.\\
Cognitive functions are attributed to cortical regions such as the frontal lobe, parietal lobe, insula, and temporal lobe. These areas play a pivotal role in various cognitive processes. On the other hand, subcortical regions, which include the basal ganglia, putamen, pallidum, thalamus, hippocampus, and brainstem, are responsible for regulating essential physiological and cognitive functions. The depiction of lesion distribution across different regions within the cortical and subcortical areas is visually presented in Fig. 2.

\begin{figure*}
\centerline{\includegraphics[width=\columnwidth]{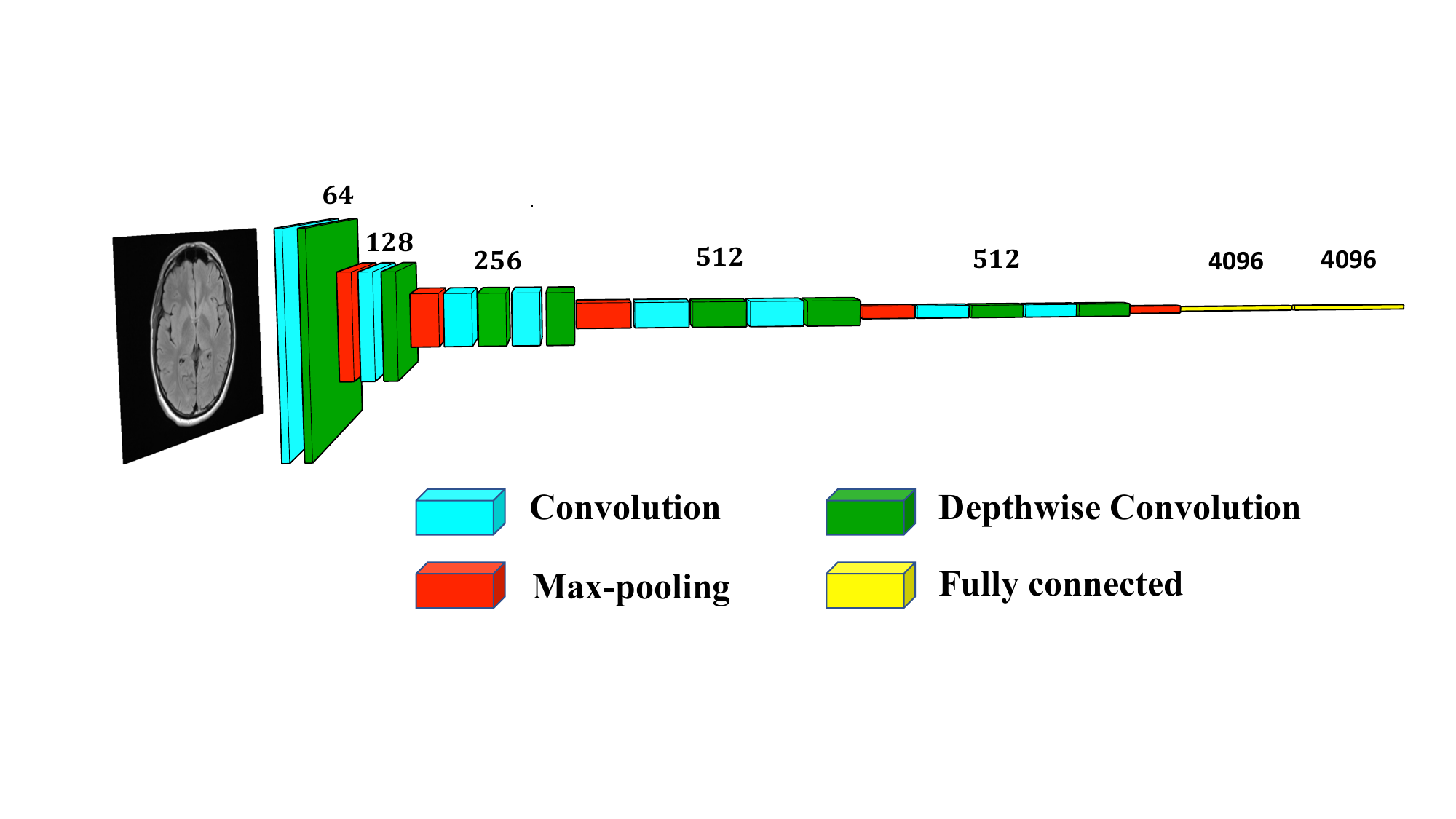}}
\caption{The Proposed Classification Model}
\label{fig1}
\end{figure*}

\section{PROPOSED METHOD}
This section provides a comprehensive introduction to the proposed methodology, which consists of classification, segmentation, aggregation functions, the integration of classification and segmentation with the aggregation function, and the combination of classification and segmentation that act as the classification. The subsection on classification and segmentation presents detailed frameworks, third subsection describes and illustrates the integrated framework that combines all the aforementioned subsections, while the final subsection describe the model when acts as a classification.
\subsection{Classification}
Medical image analysis involves a crucial task of classification, which is integral to several clinical applications. The process of classification entails labeling an image or a region of interest within an image with one or multiple class labels. This is based on particular visual characteristics or quantitative measurements. The primary purpose of medical image classification is to enhance diagnostic precision and treatment planning by automatically detecting patterns and irregularities that are symptomatic of a particular ailment or medical condition. The fundamental architecture of the network comprises an improved and modified version of the VGG model. VGG is abbreviated for Visual Geometry Group. The VGG architecture is distinguished by its uniformity and simplicity, consisting of numerous layers of 3x3 convolutional filters followed by max pooling layers and is concluded by a fully connected layer that generates the classification output. As illustrated in Fig. 3., the proposed classification model is composed of multiple layers of convolution and depthwise convolution filters, followed by max pooling. With the reduction of spatial resolution, the number of filters in each layer increases, allowing the network to capture increasingly intricate and abstract visual characteristics.

.

\begin{figure}[!t]
\centerline{\includegraphics[width=\columnwidth]{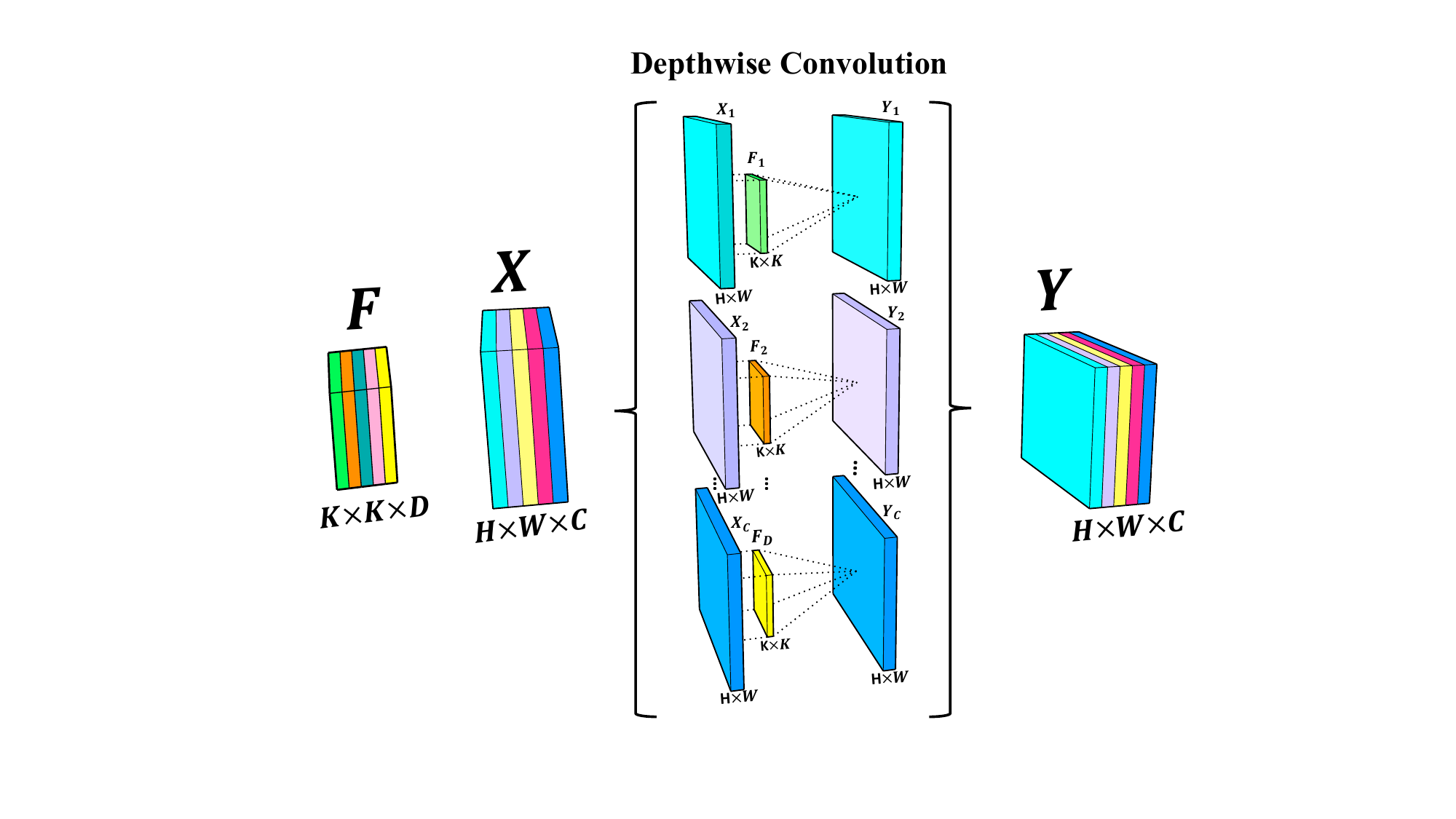}}
\caption{Depthwise Convolution }
\label{fig1}
\end{figure}

\subsubsection{Depthwise Convolution Layer}
Depthwise convolution is a common convolutional operation utilized in deep learning for image processing applications. It is a variation of the standard convolution operation that conducts a separate convolution on each channel of the input, instead of convolving all channels together using a single set of filters. This technique can notably decrease the computational cost of the convolution operation and enhance the model's efficiency.
Fig. 4. provides insight into the mathematical definition of depthwise convolution. The input tensor, X, has dimensions of H×W×C, where H×W represents the feature dimensions of height and width, and C represents the channel of the feature map. The depthwise convolution filter is defined as F with a size of K×K×C, where K represents the size of kernel windows. Depthwise convolution applies a filter to each input channel individually, producing an output tensor with the same number of channels as the input tensor. In other words, the convolution is executed between each input tensor and depthwise filter element with the same index, resulting in an output according to the following formula.

\begin{figure*}
\centerline{\includegraphics[width=\columnwidth]{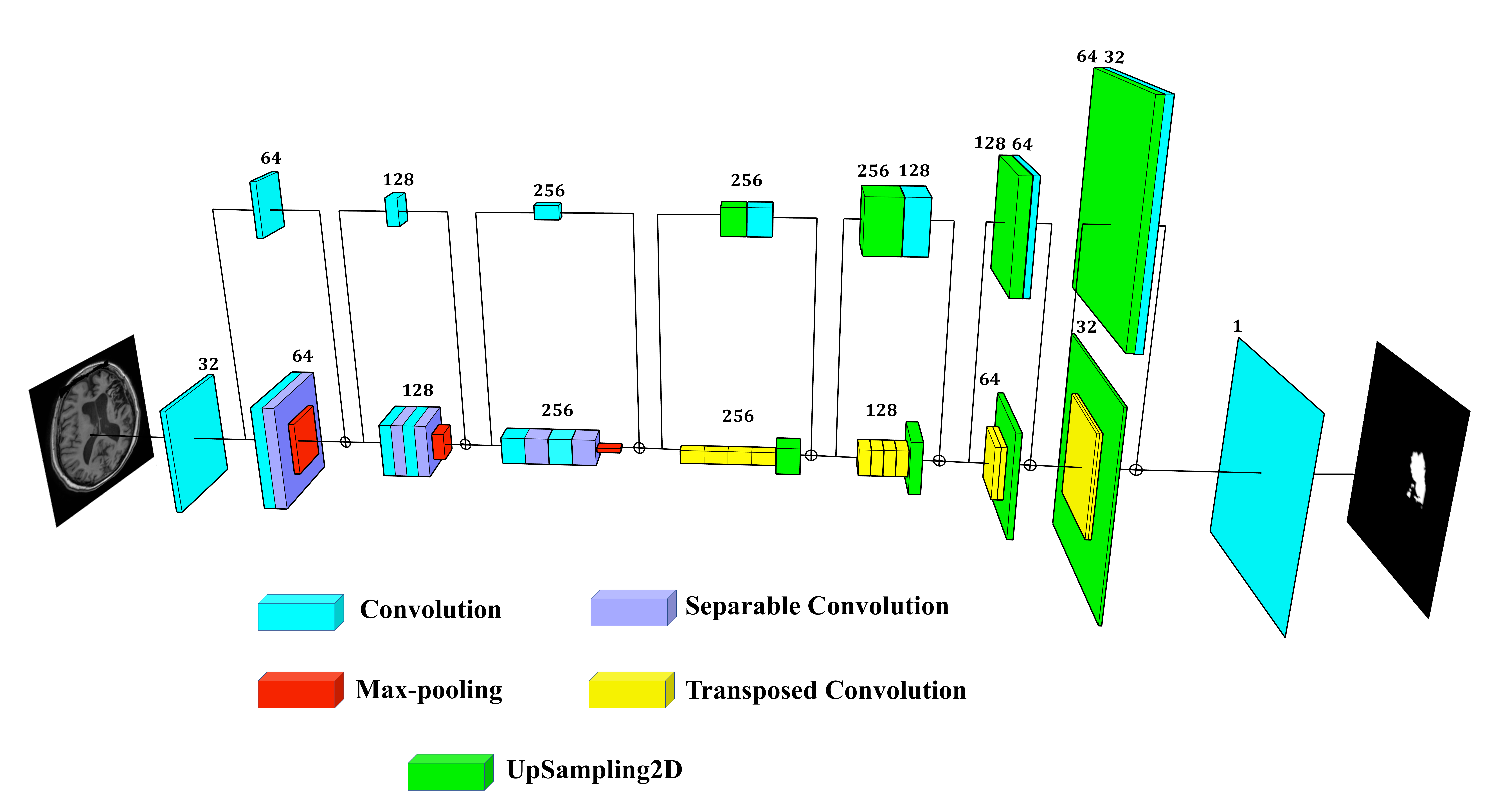}}
\caption{The Proposed Segmentation Model}
\label{fig1}
\end{figure*}

\begin{multline}
Y_n=\sum_{H_i=0}^H \sum_{W_i=0}^W Y_n\left[H_i, W_i, C_n\right]\\
Y_n\left[H_i, W_i, C_n\right]=\sum_{K_i=0}^K \sum_{K_j=0}^K  \\ X_n\left[H_i+K_i, W_i+K_j, C_n\right] * F_n\left[K_i, K_j, D_n\right] 
\end{multline}

In the equation presented, $n \in (1, 2, 3, \dots, C)$, $Y_n$ represents the n-th output channel, $X_n$ represents the n-th input channel, and $F_n$ represents the n-th depthwise convolution channel. $H_i$ and $W_i$ denote the spatial coordinates of the inputs, while $C_n$ is the index of the input channel. $K_i$ and $K_j$ represent the spatial coordinates of the depthwise convolution window, and $D_i$ represents the index of the depthwise convolution channel.

After all $Y_n$ values are generated from the convolution process, they are concatenated along the channel axis to produce the output tensor. In other words, the output tensor is formed by combining all $Y_i$ values into a single tensor, where the channel axis contains the concatenated values.

\begin{equation}
Y=\operatorname{Concat}\left(Y_1, Y_2, \ldots, Y_n\right)
\end{equation}

$Y$ represents the output of the depthwise convolution layer.\\
Depthwise convolution offers a significant reduction in parameters compared to traditional convolutional layers, making it ideal for resource-constrained devices in medical image processing. It operates across spatial dimensions while preserving the channel dimension, improving the capture of local spatial information and optimizing network capacity utilization. Depthwise convolution enables the extraction of diverse and specialized features by independently focusing on each channel, enhancing the network's representation power. On the other hand, traditional convolution incorporates cross-channel interactions, allowing for the integration of information from multiple channels and extraction of complex features. By combining depthwise and traditional convolutions sequentially, the proposed method effectively captures both local and global aspects of the data while conserving computational resources. Depthwise convolution is particularly useful in real-time applications with minimal processing delays and large trainable parameter models, reducing latency and computational overhead.

\subsection{segmentation}
Medical image segmentation is crucial for analyzing and understanding medical images by identifying and isolating different structures within the image. The network architecture consists of convolution, separable convolution, upsampling, and transposed convolution layers. It follows a contracting path with downsampling operations to capture low-level features and an expanding path with upsampling operations to reconstruct the input image using high-level features. The architecture comprises a downsampling encoder and an upsampling decoder. The encoder consists of multiple blocks of convolutional and separable convolutional layers, increasing the depth of features while decreasing the spatial dimensions. The upsampling decoder utilizes transposed convolutional layers to increase the spatial dimensions of the input, reconstructing the compressed representation obtained from the encoder. Residual connections are included in the decoder to preserve fine-grained details of the original image and improve segmentation accuracy by mitigating spatial information loss during downsampling. Overall, this architecture enables effective medical image segmentation by capturing both low-level and high-level features while preserving important spatial details through upsampling and residual connections. The proposed segmentation model is illustrated in Fig. 5.

\subsubsection{Seperable Convolution Layer}
In this architecture, as shown in Fig. 6., Separable convolution is a combination of depthwise convolution and pointwise convolution. Pointwise convolution is a type of convolutional operation used in deep learning architectures for image processing tasks. It involves applying a 1x1 convolutional kernel to each pixel of an input feature map, which results in a set of output feature maps. This process reduces the number of parameters and computational costs associated with the layer. The inclusion of separable convolutional layers in this architecture results in an improved feature extraction capability for the encoder, with a simultaneous reduction in computational complexity.\\
According to Fig. 6., the separable convolution is composed of two components: the depthwise convolution and the pointwise convolution. The depthwise component is computed using the formula provided, where $Y'$ denotes the outcome of the convolution of the input and the depthwise filter.

\begin{figure}[!t]
\centerline{\includegraphics[width=\columnwidth]{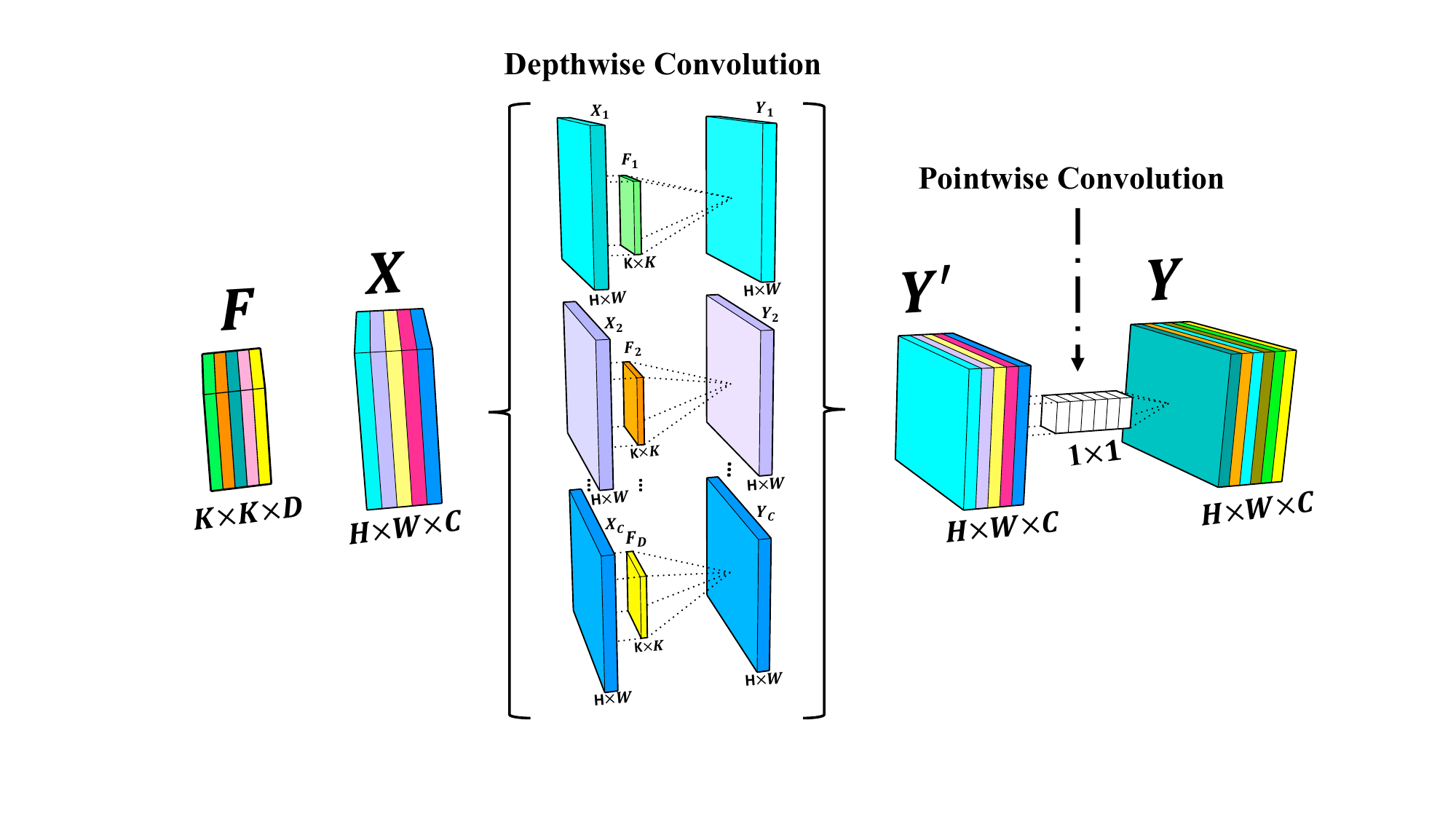}}
\caption{Separable Convolution Architecture}
\label{fig1}
\end{figure}

\begin{multline}
Y_n^{\prime}=\sum_{H_i=0}^H \sum_{W_i=0}^W Y_n^{\prime}\left[H_i, W_i, C_n\right] \\
Y_n^{\prime}\left[H_i, W_i, C_n\right]=\sum_{K_i=0}^K \sum_{K_j=0}^K \\ X_n\left[H_i+K_i, W_i+K_j, C_n\right] * F_n\left[K_i, K_j, D_n\right] \\ \\
Y^{\prime}=\operatorname{Concat}\left(Y_1^{\prime}, Y_2^{\prime}, \ldots, Y_n^{\prime}\right)\\
\end{multline}

In the pointwise convolution step, the result Y' is convolved with a pointwise filter of size 1×1×C×$C'$, where c represents the number of channels of the feature maps and $c'$ denotes the number of pointwise filters. The pointwise filter can be represented as a 4D tensor of size 1×1×c×$c'$, where each element in the filter corresponds to a scalar weight used to compute the output value.As a result, the output is obtained using the formula provided below.

\begin{multline}
Y_m =\sum_{H_i=0}^H \sum_{W_i=0}^W Y_m\left[H_i, W_i, C_m^{\prime}\right] \\
Y_m\left[H_i, W_i, C_m^{\prime}\right]=\sum_{C_n=0}^{c_n} Y^{\prime}\left[H_i, W_i, C_n\right] * P\left[1,1, C_n, C_m^{\prime}\right]\\ \\
Y =\operatorname{Concat}\left(Y_1, Y_1, \ldots, Y_m\right)\\
\end{multline}

The pointwise filter P is a 4D tensor with the element P[1, 1, C, $C'$] denoting the value of the 1x1 convolutional kernel at position (1, 1, c, $C'$). The index $C'$ represents the m-th element in the set of pointwise filters, where $m \in (1, 2, 3, \ldots, C')$. \\


The separable convolution layer is beneficial for capturing both local and cross-level features by leveraging spatial and channel-wise interactions effectively. It is particularly useful in models that require a larger receptive field and the extraction of both local and global features. In contrast to traditional convolution, which considers spatial and channel-wise interactions concurrently, separable convolution separates the operation into depthwise convolution and pointwise convolution. Depthwise convolution captures spatial information independently for each input channel, while pointwise convolution combines the output channels from depthwise convolution to generate the final output. This separation allows for computational efficiency. Although depthwise convolution in separable convolution can capture local features, the subsequent pointwise convolution enables the capture of cross-level features by aggregating information across channels.However, it's important to note that while separable convolution can capture cross-level features, its expressiveness may be somewhat limited compared to traditional convolution due to the absence of direct simultaneous spatial and channel-wise interactions.

\begin{figure*}
\centerline{\includegraphics[width=\columnwidth]{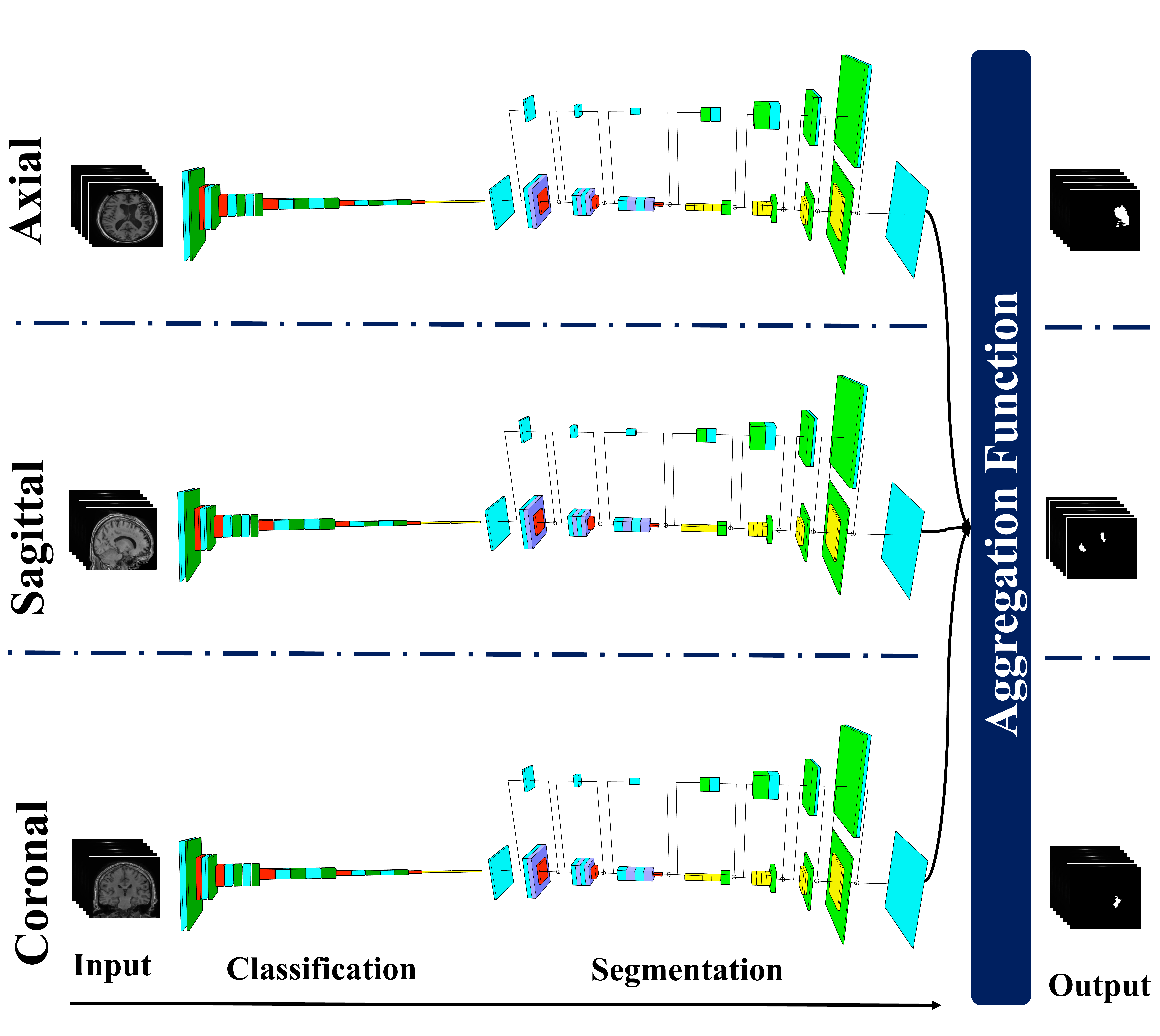}}
\caption{VRU-net: The Overall Of The Proposed Model}
\label{fig1}
\end{figure*}

\subsection{Three planes for Segmentation}

Magnetic resonance imaging (MRI) images of the brain consist of three-dimensional tensor planes: Axial, Sagittal, and Coronal. Each plane provides a different perspective of the brain's structure. The Axial Plane divides the body into superior and inferior sections, with the x-axis representing left-right and the y-axis representing anterior-posterior directions. The Sagittal Plane partitions the body into left and right portions along the mid-sagittal plane, with the y-axis representing anterior-posterior and the z-axis representing superior-inferior directions. The Coronal Plane separates the body into anterior and posterior sections, with the x-axis representing left-right and the z-axis representing superior-inferior directions. For segmentation tasks in medical imaging, both three-dimensional (3D) and two-dimensional (2D) models are used, with differences in dimensionality, data representation, complexity, and training data requirements. The choice of model depends on the specific task, data availability, and the importance of spatial information. While 3D models are effective, they can be time-consuming and require significant memory and processing capabilities. To address these challenges, researchers have developed models trained on single planes. In brain segmentation, the Axial plane is commonly used, but relying solely on this plane may overlook important information. Lesions may have unclear boundaries in the Axial plane but clearer boundaries in the Sagittal and Coronal planes. Therefore, incorporating all three planes can improve accuracy in lesion segmentation. The proposed segmentation approach involves leveraging expertise in the Axial, Sagittal, and Coronal planes by utilizing an aggregation function. By considering multiple planes, the model can accurately identify lesions with well-defined boundaries and achieve precise segmentations, particularly for ischemic stroke segmentation.

The proposed segmentation approach revolves around the inclusion of these three planes and is implemented through the utilization of an aggregation function.

\subsection{Aggregation Function}
he MRI images are represented as a three-dimensional tensor, denoted as $M \in \mathbb{R}^{197 \times 233 \times 189}$, wherein each MRI consists of three distinct planes: Axial, Sagittal, and Coronal. These planes are represented as $A_{x-y}^M \in \mathbb{R}^{197 \times 233}, S_{y-z}^M \in \mathbb{R}^{233 \times 189},$ and $C_{x-z}^M \in \mathbb{R}^{197 \times 189}$, respectively. The intersection of these three planes, denoted as $A_{x-y}^M \cap S_{y-z}^M \cap C_{x-z}^M$ corresponds to the entire MRI image M.
For each plane, the segmentation model is represented as $f_{x-y}, f_{y-z}$, and $f_{x-z}$ . The final result, denoted as $P^M$ , is obtained through an aggregation function, F, which takes the outputs of the segmentation models for each plane as its inputs:

\begin{equation}
P^M=F\left(f_{x-y}\left(A_{x-y}^M\right), f_{y-z}\left(S_{y-z}^M\right), f_{x-z}\left(C_{x-z}^M\right)\right)
\end{equation}

Here, $P^M$  represents the predicted probability of the single MRI image, and the specific choice of aggregation function depends on the desired outcome. In the proposed method, the objective of the aggregation function is to reduce false positive values in segmentation and enhance the segmentation performance in each plane by leveraging information from the other planes.\\
Thus, the most suitable aggregation function in this context is defined as follows: for each pixel in each plane that the segmentation model identifies as a lesion, if the segmentation model in the other planes also identifies that pixel as a lesion, it is considered a lesion. Otherwise, the pixel is classified as part of the healthy brain regions.
To compute $P^M$, we calculate the predicted probabilities for each plane individually:

\begin{equation}
\begin{aligned}
& \hat{p}_{x-y}^M=f_{x-y}\left(A_{x-y}^M\right) \\
& \hat{p}_{y-z}^M=f_{y-z}\left(S_{y-z}^M\right) \\
& \hat{p}_{x-z}^M=f_{x-z}\left(C_{x-z}^M\right)
\end{aligned}
\end{equation}

Finally, the aggregation function is applied as:

\begin{multline}
P^M=F\left(f_{x-y}\left(A_{x-y}^M\right), f_{y-z}\left(S_{y-z}^M\right), f_{x-z}\left(C_{x-z}^M\right)\right)=\\
F\left(\hat{p}_{x-y}^M, \hat{p}_{y-z}^M, \hat{p}_{x-z}^M\right)=T\left(\hat{p}_{x-y}^M+\hat{p}_{y-z}^M+\hat{p}_{x-z}^M\right)
\end{multline}

Here, T represents the threshold value, which is set to 3. The aggregation function accepts a pixel as a lesion only if all three segmentation models predict it as a lesion, resulting in a sum of 3 for that pixel.

\subsection{Visual-Residual U-net}
The proposed methodology focuses on analyzing 3D MRI data by decomposing it into 2D image slices representing the axial, sagittal, and coronal planes. The model architecture consists of classification, segmentation, and aggregation components. Six models are trained, two for each plane, to sequentially process the image slices. The classification model predicts the presence of a stroke lesion, and only those slices designated as lesions proceed to the segmentation model. If no lesion is identified, the segmented images consist of zero values. The output from each model is then aggregated to determine the final segmented result. The model architecture, depicted in Fig. 7., comprises three primary components: classification, segmentation, and an aggregation function.

\subsection{Act as a Classification}
The proposed method aims to accurately segment stroke lesions in MRI images. Only image slices classified as lesion classes by the classification model proceed to the segmentation model. However, there are cases where the classification model predicts the presence of a lesion in slices without actual lesions. These false positive predictions are used as an indication to evaluate and improve the accuracy of the classification model. The segmentation model refrained from segmenting any regions of the images that the classification process identified as a lesion, despite the absence of a genuine lesion in the slice.
On the other hand, slices containing lesions but predicted as non-lesion classes are not processed further, resulting in false negative predictions that are not addressed in this process.

\section{Implementation}

The proposed methodology has been tested on the NVIDIA GeForce GTX 1080-Ti GPU and implemented using the Keras and TensorFlow frameworks. In the classification phase, the Adam optimizer with a learning rate of 0.00001 is used. Max-pooling is performed with a size of 2 and stride of 2, and both the depthwise convolution and convolution layers have a kernel size and stride of 1. To prevent overfitting, a callback function is used to monitor the model's performance on test metrics and terminate training if there is no improvement. The validation loss is the metric tracked by the callback function. In the classification mode, a batch size of 16 is used, and the loss function is binary cross-entropy with the activation function being Rectified Linear Unit (RELU). In the segmentation phase, max-pooling is performed with a size of 3 and a stride of 2. The segmentation model uses binary cross-entropy as the loss function and employs the Adam optimizer with a learning rate of 0.001. A batch size of 16 is also used in the segmentation phase.

\subsection{Data division}

This article focuses on utilizing the ATLAS database, which consists of 655 3D cases represented by axial, sagittal, and coronal planes. The methodology involves two separate networks: a classification network and a segmentation network, both trained on 2D slices. The classification model is trained on both lesion and normal slices, while the segmentation model is trained solely on lesion slices. The data is divided into two steps for training and evaluation purposes. In the first step, 20\% of the total cases (131 cases) are used to evaluate the model in 3D mode on a per-case basis. The remaining 80\% (524 cases) are used for training and testing the classification and segmentation models in 2D mode, the more detail is reported in Table 3. Data balancing is crucial for reliable evaluation in the classification training, while in the segmentation procedure, data balancing is not necessary as the model only trains on lesion data. The number of samples required for both classification and segmentation tasks is defined in Table 4, considering the greater number of lesion slices compared to normal slices and the importance of balanced data for classification.

\begin{table}

\centering
\caption{Number of Cases, Total Slices, Total Lesion Slices, and Total Normal Slices are reported ( Total cases are 655 and 524 Cases are used for Train and Test data for 2D slices, the rest cases used for 3D testing}
\label{table}

\setlength{\tabcolsep}{0.4pt}

\begin{tabular}{c p{0.2\linewidth} c c c }
\toprule
 \textbf{Cases} & $\begin{array}{c}\textbf { Modality } \end{array}$ & \textbf{Total Slices} & $\begin{array}{c}\textbf { Total lesion } \end{array}$ & $\begin{array}{c}\textbf { Total Normal } \end{array}$ \\

\hline \multirow{3}{*}{ \textbf{655 cases} } & \centering Axial & 123,795 & 32,989 & 90806 \\
 & \centering Sagittal & 123,795 & 23,455 & 100,340 \\
 & \centering Coronal & 123,795 & 33,919 & 89,876 \\
 
\hline \multirow{3}{*}{ \textbf{524 cases} } & \centering Axial & 99,036 & 25,895 & 73,141 \\
 & \centering Sagittal & 99,036 & 18,259 & 80,777 \\
 
 & \centering Coronal & 99,036 & 26,705 & 72,331 \\
\bottomrule
\end{tabular}
\label{tab3}
\end{table}

\begin{table}
\centering
\caption{The number of slices for classification and segmentation model - the number of lesion and normal slices are equal for classification task - for segmentation model only lesion slices are used}
\label{table}
\setlength{\tabcolsep}{0.3pt}
\begin{tabular}{p{0.2\linewidth} p{0.2\linewidth} c c c c }
\toprule
\textbf{Modality} & \textbf{Task} & $\begin{array}{c}\textbf { Train } \\
\textbf { Slices }\end{array}$ & $\begin{array}{c}\textbf { Test }\\ 
\textbf { Slices }\end{array}$ & $\begin{array}{c}\textbf { Lesion } \\
\textbf { Slices } \\
\textbf { (Train) }\end{array}$ & $\begin{array}{c}\textbf { Lesion } \\
\textbf { Slices } \\
\textbf { (Test) }\end{array}$ \\
\hline \multirow{2}{*}{ \textbf{Axial} } & \textbf{Classification} & 41,432 & 10,358 & 20,716 & 5,179 \\
 & \textbf{Segmentation} & 20,716 & 5,179 & 20,716 & 5,179 \\
\hline \multirow{2}{*}{ \textbf{Sagittal} } & \textbf{Classification} & 29,214 & 7,304 & 14,607 & 3,652 \\
 & \textbf{Segmentation} & 14,607 & 3,652 & 14,607 & 3,652 \\
\hline \multirow{2}{*}{ \textbf{Coronal} } & \textbf{Classification} & 42,728 & 10,682 & 21,364 & 5,341 \\
 & \textbf{Segmentation} & 21,364 & 5,341 & 21,364 & 5,341 \\
\bottomrule
\end{tabular}
\label{tab3}
\end{table}

\subsection{metrics}

The proposed technique incorporates both classification and segmentation processes for evaluating performance using various metrics. For classification, metrics such as precision, recall, F1-score, accuracy, sensitivity, and specificity are utilized. In segmentation, the dice similarity coefficient, precision, and recall are used for evaluation. Due to imbalances between foreground and background in medical image segmentation, accuracy alone is an unreliable metric. The metrics are defined by four terms: true positives (TP), true negatives (TN), false positives (FP), and false negatives (FN). True positives are the pixels accurately predicted as belonging to the positive class, while true negatives are pixels correctly identified as belonging to the negative class. False positives and false negatives represent pixels erroneously classified as positive and negative, respectively. More detailed explanation of these metrics and their application to classification and segmentation can be found in the following subsection.

\subsubsection{Classification}
The performance of the model is evaluated using various metrics, including precision, recall, F1-score, accuracy, sensitivity, and specificity. These metrics serve as indicators of the model's ability to accurately predict positive and negative classes and determine its overall performance.
\textbf{Precision}
Precision is a metric that calculates the quantification of the model in the number of correct positive predictions and divides the true positive into all the positive predictions. The value of this metric is between 0 and 1. The 1 reveals the best results. The formulation is : 
Precision is calculated by dividing the true positives by anything predicted as a positive.

\begin{equation}
\text { Precision }=\frac{T P}{T P+F P}
\end{equation}

\textbf{Recall}

The recall is the ratio of the true positive predictions to the total number of instances that should have been identified as positive

\begin{equation}
\text { Recall }=\frac{T P}{T P+F N}
\end{equation}

\textbf{F1-score}

The F1 score is a single metric that combines both recall and precision. A high F1 score indicates good performance, whereas a low score indicates poor performance. In the given statement, the model is characterized as having excellent recall but poor precision.

\begin{equation}
F_{-} \text {measure }=2 . \frac{\text { Precision } \times Recall}{\text { Precision }+Recall}
\end{equation}

\textbf{Sensitivity}

In the context of machine learning, this metric assesses the model's ability to accurately predict the true positive instances for each class.

\begin{equation}
\text { Sensitivity }=\frac{T P}{T P+F P}
\end{equation}

\textbf{Specificity}

This metric reveals the ability of the model to predict the true negative of each class.

\begin{equation}
\text { Specificity }=\frac{T N}{T N+F N}
\end{equation}

\textbf{Accuracy}

These metrics quantify the performance of a model in terms of correct predictions.

\begin{equation}
\text { Accuracy }=\frac{T P+T N}{T P+T N+F P+F N}
\end{equation}

\subsubsection{Segmentation}

In the case of MRI images, the masked label is represented in binary format, with grayscale values of 1 and 0 corresponding to the lesion and non-lesion regions of the brain, respectively.

\textbf{Dice Score}

The degree of overlap between two binary images can be quantified using the Dice score, which is expressed mathematically as follows:

\begin{equation}
D S C=\frac{2 T P}{F P+2 T P+F N}
\end{equation}

The Dice score is a metric that ranges from 0 to 1, with a score of 0 indicating no overlap and a score of 1 indicating complete overlap between two binary images.

\section{Experimental and result}

The study presents a comprehensive methodology consisting of multiple components, each evaluated independently to assess their performance. The evaluation is conducted systematically, comparing the proposed approach with other state-of-the-art methods.
In the first subsection, the evaluation metrics for the classification model are reported, comparing them to the VGG model in terms of performance and the number of trainable parameters. This comparison provides a thorough assessment of the classification model's effectiveness. Moving to the second subsection, the evaluation metrics for the segmentation model are presented, along with a comparison of parameters and performance against other approaches. This evaluation examines the segmentation model's ability to accurately segment desired regions. The third subsection focuses on evaluating the segmentation model combined with an aggregation function, comparing it to the segmentation model without aggregation. This analysis highlights the potential of the aggregation function in improving segmentation performance and emphasizes its importance in the methodology. In the fourth subsection, the evaluation shifts to the integrated model that combines both classification and segmentation components with an aggregation function. The performance of this integrated model is assessed and compared to other approaches, providing a comprehensive understanding of its effectiveness. Lastly, the final subsection explores the model's performance when considering the classification and segmentation models as a unified classification algorithm. This evaluation offers insights into the overall performance of the model and its ability to handle both classification and segmentation tasks effectively.

\begin{figure*}[!t]
\centerline{\includegraphics[width=\columnwidth]{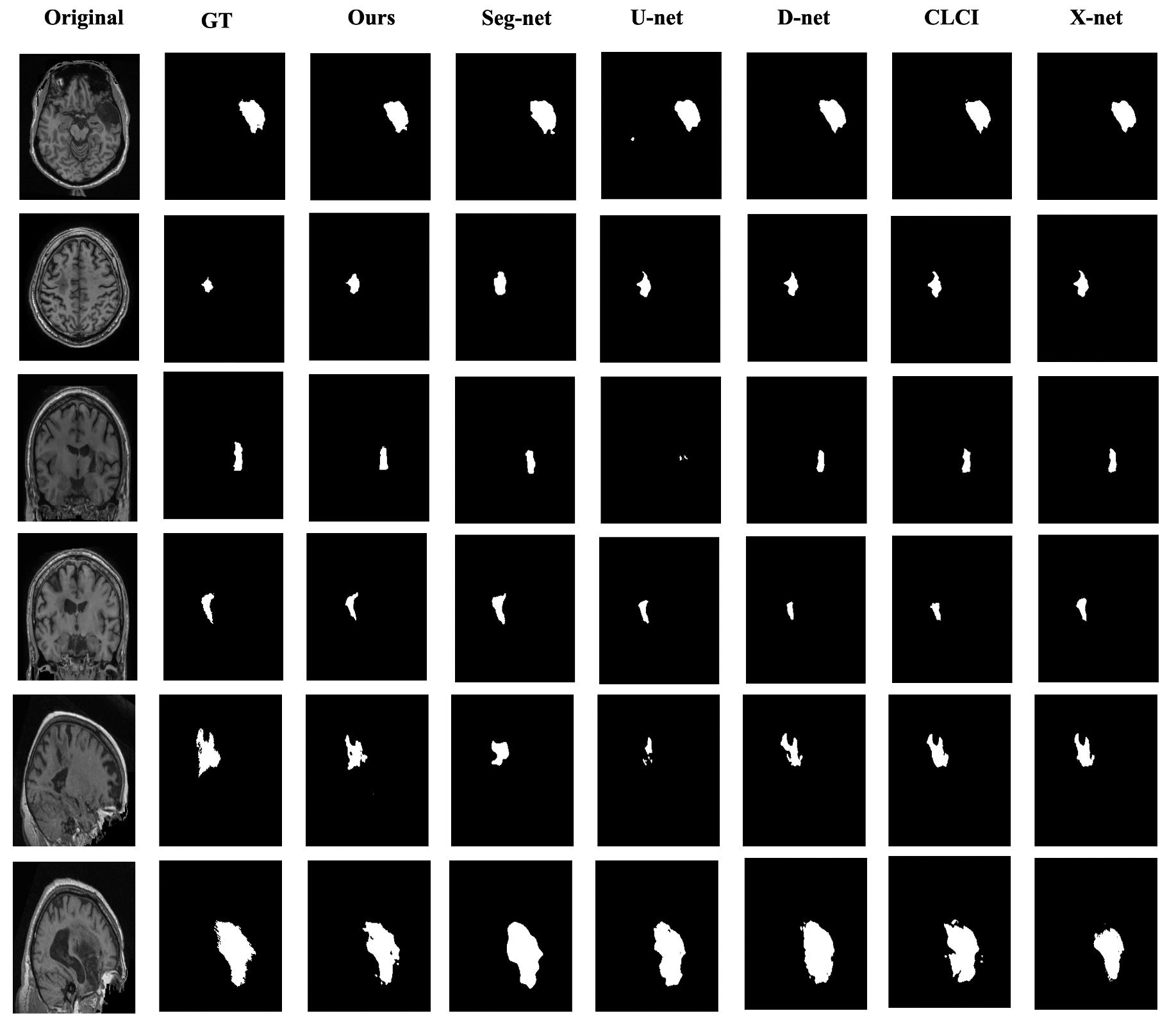}}
\caption{Magnetization as a function of applied field.
It is good practice to explain the significance of the figure in the caption.}
\label{fig1}
\end{figure*}

\begin{table}
\centering
\caption{The Report on the Proposed Classification Model and Its Comparison to VGG-19 }
\label{table}
\setlength{\tabcolsep}{0.3pt}
\begin{tabular}{ p{0.1\linewidth} p{0.11\linewidth} p{0.1\linewidth} p{0.1\linewidth} p{0.1\linewidth} p{0.1\linewidth} p{0.1\linewidth} p{0.1\linewidth} p{0.20\linewidth} }

\hline \centering $\begin{array}{c}\\ \textbf{Mod} \\
\text {}\end{array}$ & \centering \textbf{Methods} & \centering \textbf{Pre.} & \centering \textbf{Rec.} & \centering \textbf{F1} & \centering \textbf{Acc.} & \centering \textbf{Sen.} & $\begin{array}{c} \\ \textbf{Spe.} \\ \text {}\end{array}$ & $\begin{array}{c} \\ \textbf{Parameters} \\ \text {}\end{array}$\\



\hline \multirow{2}{*}{\centering \raisebox{-1.0\height}{\rotatebox{90}{\textbf{\large Axial}}}} & \centering $\begin{array}{c} \\ \textbf {VGG} \\
\text {}\end{array}$ & \centering 95.35 & \centering 95.35 & \centering 95.35 & \centering 95.35 & \centering 95.40 & \centering 95.28 & $\begin{array}{c} \\ \text {$124,893,121$ } \\
\text {}\end{array}$ 
 \\
 & \centering \textbf{OURS} & \centering 92.88 & \centering 92.88 & \centering 92.88 & \centering 92.88 & \centering 92.68 & \centering 93.06 & $\begin{array}{c} \\ \text {$105,913,217$ } \\
\text {}\end{array}$ 
 \\

\hline \multirow{2}{*}{\centering \raisebox{-0.7\height}{\rotatebox{90}{\textbf{\large Sagittal}}}} & \centering $\begin{array}{c} \\ \textbf {VGG} \\
\text {}\end{array}$ & \centering 93.32 & \centering 93.29 & \centering 93.29 & \centering 93.29 & \centering 94.63 & \centering 91.95 & $\begin{array}{c} \\ \text {$110,213,057$ } \\
\text {}\end{array}$ 
 \\
 & \centering $\begin{array}{c} \\ \textbf {OURS} \\
\text {}\end{array}$ & \centering 91.25 & \centering 91.24 & \centering 91.24 & \centering 91.24 & \centering 91.86 & \centering 90.61 & $\begin{array}{c} \\ \text {$91,821,697$ } \\
\text {}\end{array}$
 \\
\hline \multirow{2}{*}{\centering \raisebox{-0.7\height}{\rotatebox{90}{\textbf{\large Coronal}}}}& \centering $\begin{array}{c} \\ \textbf {VGG} \\
\text {}\end{array}$ & \centering 95.31 & \centering 95.31 & \centering 95.31 & \centering 95.31 & \centering 96.59 & \centering 96.03 &  $\begin{array}{c} \\ \text {$99,727,297$ } \\
\text {}\end{array}$
 \\
 & \centering $\begin{array}{c} \\ \textbf {OURS} \\
\text {}\end{array}$ & \centering 92.01 & \centering 91.98 & \centering 91.98 & \centering 91.98 & \centering 93.35 & \centering 90.60 & $\begin{array}{c} \\ \text {$80,747,393$ } \\
\text {}\end{array}$
 \\
\hline
\end{tabular}
\begin{tablenotes}
\item \text{\textbf{Mod:} Modality; \textbf{Pre:} Precision; \textbf{Rec} Recall;
\textbf{F1:} F1-score; \textbf{Acc:} Accuracy}
\end{tablenotes}
\begin{tablenotes}
\item \text{ \textbf{Sen:} Sensitivity; \textbf{Spe:} Specificity}
\end{tablenotes}

\label{tab3}
\end{table}

\subsection{classification}
The study proposes an improved version of the VGG model for classification, incorporating a depthwise convolution layer as a key enhancement. This modification reduces the number of trainable parameters and computational complexity. The performance of the classification model is evaluated using various metrics, and the results are summarized in Table 5. Additionally, Table 5 presents a comprehensive comparison of the classification model's performance with the conventional VGG model. Table 3 provides a detailed analysis of the number of parameters utilized by both models. It is noteworthy that the performance evaluation and parameter analysis are conducted separately for each anatomical plane, ensuring a thorough assessment.

\subsection{segmentation}

The segmentation model in this study utilizes convolution and separable convolution layers, with the latter contributing to a reduction in trainable parameters and faster training. The performance of the model in segmenting lesion stroke is presented in Table 6, which includes evaluation metrics specific to medical image segmentation. In Table 6, a comparison is made between the proposed model and a state-of-the-art approach from previous years for ischemic segmentation. Additionally, Table 6 provides a comparison of the number of parameters between the proposed model and other state-of-the-art models. It should be noted that the comparison is based on lesion data exclusively, consistent with the other approaches. The proposed segmentation model exhibits a significantly lower number of trainable parameters, as evidenced by the data presented in Table 10, in comparison to other state-of-the-art models.

\begin{table}
\centering
\caption{Comparison of Trainable Parameters in the Proposed Segmentation Model and Other State-of-the-Art Models}
\label{table}
\setlength{\tabcolsep}{0.3pt}

\begin{tabular}{ p{0.20\linewidth} c }
\hline \textbf{Model} & \textbf{Number of trainable parameters} \\
\hline \textbf{X-net} & $\mathbf{1 5 , 1 1 7 , 6 4 6}$ \\
\hline \textbf{CLCI} & $\mathbf{3 6 , 8 1 3 , 0 2 5}$ \\
\hline \textbf{D-net} & $\mathbf{8 , 6 3 3 , 3 7 9}$ \\
\hline \textbf{Seg-net} & $\mathbf{1 0 , 1 8 3 , 9 3 7}$ \\
\hline \textbf{u-net} & $\mathbf{3 1 , 0 3 0 , 5 9 3}$ \\
\hline \textbf{OURS} & $\mathbf{4 , 6 9 7 , 9 2 1 }$ \\
\hline
\end{tabular}

\label{tab3}
\end{table}

\begin{table}
\centering
\caption{The Report on the Proposed Segmentation Model and Its Comparative Analysis with Other State-of-the-Art Models in Ischemic Lesion Segmentation, All Trained Exclusively on Lesion Slices}
\label{table}
\setlength{\tabcolsep}{0.3pt}

\begin{tabular}{p{0.10\linewidth} c c c c }
\hline \textbf {Mod} & \textbf {Method} & \textbf {DICE} & \textbf {precision} & \textbf {recall} \\
\hline \\ \multirow{6}{*}{\centering \raisebox{-1.9\height}{\rotatebox{90}{\textbf{\large AXIAL}}}} & \textbf{X-net} & $0.4740 \pm 0.3275$ & \textbf{ }$0.5909 \pm 0.4186$ \textbf{ } & \textbf{  }$0.4966 \pm 0.3912$ \\ \\

 & \textbf{CLCI}  & $0.7291 \pm 0.2403$ & $0.7970 \pm 0.2851$ & $0.7287 \pm 0.2950$ \\ \\
 & \textbf{D-net}   & $0.6850 \pm 0.2602$ & $0.7680 \pm 0.3213$ & $0.6823 \pm 0.3255$ \\ \\
 & \textbf{Seg-Net} & $0.7314 \pm 0.2395$ & $0.7963 \pm 0.2712$ & $0.7310 \pm 0.2897$ \\ \\
 & \textbf{U-Net} & $0.5514 \pm 0.3142$ & $0.6390 \pm 0.3843$ & $0.5616 \pm 0.3742$ \\ \\
& \textbf{Ours} & $\mathbf{0.7674 \pm 0.2154}$ & $\mathbf{0.8117 \pm 0.2434}$ & $\mathbf{0.7768 \pm 0.2598}$ \\ \\
\hline \\ \multirow{6}{*}{\centering \raisebox{-1.4\height}{\rotatebox{90}{\textbf{\large SAGGITAL}}}} & \textbf{X-net} & $0.43552 \pm 0.3180$ & $0.5514 \pm 0.3953$ & $0.5256 \pm 0.3905$ \\ \\
& \textbf{CLCI}  & $0.5839 \pm 0.2999$ & $0.6694 \pm 0.3616$ & $0.6106 \pm 0.3620$ \\ \\
& \textbf{D-net} & $0.4483 \pm 0.3168$ & $0.5686 \pm 0.4135$ & $0.4664 \pm 0.3888$ \\ \\
& \textbf{Seg-Net} & $0.5768 \pm 0.2977$ & $0.6586 \pm 0.3406$ & $0.6419 \pm 0.3578$ \\ \\
& \textbf{U-Net} & $0.5254 \pm 0.3073$ & $0.6462 \pm 0.3987$ & $0.5116 \pm 0.3646$ \\ \\
 & \textbf{Ours}  & $\mathbf{0.7081 \pm 0.2430}$ & $\mathbf{0.7827 \pm 0.2826}$ & $\mathbf{0.6997 \pm 0.2939}$ \\ \\
\hline \\ \multirow{6}{*}{\centering \raisebox{-1.4\height}{\rotatebox{90}{\textbf{\large CORONAL}}}} & \textbf{X-net}  & $0.4910 \pm 0.3074$ & $0.6192 \pm 0.4178$ & $0.5009 \pm 0.3845$ \\ \\
 & \textbf{CLCI} & $0.7248 \pm 0.2384$ & $0.7755 \pm 0.2670$ & $0.7796 \pm 0.2749$ \\ \\
 & \textbf{D-net} & $0.6958 \pm 0.2526$ & $0.7835 \pm 0.2994$ & $0.7000 \pm 0.3064$ \\ \\
& \textbf{Seg-Net} & $0.7595 \pm 0.2218$ & $0.8210 \pm 0.2379$ & $0.7829 \pm 0.2605$ \\ \\
& \textbf{U-Net} & $0.6166 \pm 0.2844$ & $0.7249 \pm 0.3483$ & $0.6360 \pm 0.3442$ \\ \\
& \textbf{Ours}  & $\mathbf{0.7725 \pm 0.2074}$ & $\mathbf{0.8234 \pm 0.2347}$ & $\mathbf{0.7823 \pm 0.2461}$ \\ \\
\hline
\end{tabular}
\begin{tablenotes}
\item \text{ \textbf{Mod:} Modality}
\end{tablenotes}
\label{tab3}
\end{table}

\begin{table}[htbp]
\centering
\caption{The Report on the Combination of the Proposed Classification and Segmentation Models and Their Comparative Analysis with Other State-of-the-Art Models in Ischemic Lesion Segmentation, Including Models Trained on Both Lesion and Normal Slices}
\label{table}
\setlength{\tabcolsep}{0.3pt}

\begin{tabular}{p{0.12\linewidth} p{0.12\linewidth} p{0.27\linewidth} p{0.26\linewidth} p{0.26\linewidth} }
\hline Mod & Method & \centering DICE & \centering \textbf{precision} & \textbf{  Recall} \\
\hline \\ \multirow{7}{*}{\centering \raisebox{-1.4\height}{\rotatebox{90}{\textbf{\large AXIAL}}}} & \textbf{X-net} & $0.3946 \pm 0.2681$ & $0.4937 \pm 0.4064$ & $0.3803 \pm 0.3178$ \\ \\
 & \textbf{CLCI} & $0.5250 \pm 0.2944$ & $0.6128 \pm 0.4128$ & $0.5586 \pm 0.3564$ \\ \\
 & \textbf{D-Net}  & $0.4989 \pm 0.3184$ & $0.5883 \pm 0.4229$ & $0.4902 \pm 0.3334$ \\ \\
 & \textbf{Seg-Net} & $0.5683 \pm 0.24617$ & $0.6176 \pm 0.3299$ & $0.5502 \pm 0.2897$ \\ \\
 & \textbf{U-Net}  & $0.4681 \pm 0.2541$ & $0.5226 \pm 0.3343$ & $0.4743 \pm 0.3252$ \\ \\
 & \textbf{OURS}  & $\mathbf{0.7696 \pm 0.2176}$ & $\mathbf{0.7766 \pm 0.2939}$ & $\mathbf{0.7469 \pm 0.3015}$ \\ \\
\hline  \\  \multirow{6}{*}{\centering \raisebox{-1.4\height}{\rotatebox{90}{\textbf{\large SAGGITAL}}}} & \textbf{X-net}  & $0.3557 \pm 0.3761$ & $0.4359 \pm 0.4001$ & $0.3843 \pm 0.2853$ \\ \\
& \textbf{CLCI}  & $0.4976 \pm 0.2474$ & $0.5461\pm 0.3347$ & $0.4817 \pm 0.3038$ \\ \\
& \textbf{D-net}  & $0.4104 \pm 0.2282$ & $0.5076 \pm 0.3181$ & $0.3735 \pm 0.2450$ \\ \\
& \textbf{Seg-Net}  & $0.4887 \pm 0.2309$ & $0.5457 \pm 0.3209$ & $0.5456 \pm 0.2505$ \\ \\
 & \textbf{U-Net}  & $0.4381 \pm 0.3529$ & $0.5226 \pm 0.3785$ & $0.4143 \pm 0.3182$ \\ \\
 & \textbf{OURS}  & $\mathbf{0.7064 \pm 0.2472}$ & $\mathbf{0.7354 \pm 0.3358}$ & $\mathbf{0.6601 \pm 0.3337}$ \\ \\
 \hline \\ \multirow{7}{*}{\centering \raisebox{-1.4\height}{\rotatebox{90}{\textbf{\large CORONAL}}}}& \textbf{X-net}   & $0.4133 \pm 0.2771$ & $0.4303 \pm 0.4209$ & $0.4241 \pm 0.3061$ \\ \\
 & \textbf{CLCI}  & $0.5366 \pm 0.4710$ & $0.3084 \pm 0.4557$ & $0.2960 \pm 0.3991$ \\ \\
 & \textbf{D-net}  & $0.4938 \pm 0.2793$ & $0.4456 \pm 0.3980$ & $0.4228 \pm 0.3405$ \\ \\
 & \textbf{Seg-Net}  & $0.5777 \pm 0.3680$ & $0.4105 \pm 0.4436$ & $0.4314 \pm 0.4326$ \\ \\
 & \textbf{U-Net}  & $0.46426 \pm 0.1966$ & $0.3627 \pm 0.3375$ & $0.3914 \pm 0.1758$ \\ \\
& \textbf{OURS} & $\mathbf{0.7666 \pm 0.2209}$ & $\mathbf{0.7651 \pm 0.3137}$ & $\mathbf{0.7287 \pm 0.3146}$ \\ \\
\hline
\end{tabular}
\begin{tablenotes}
\item \text{ \textbf{Mod:} Modality}
\end{tablenotes}
\label{tab3}
\end{table}

\begin{table*}[!t]
\centering
\caption{Segmentation Method Overview and Evaluation Results on 131 MRI 3D Images with/without aggregation function}
\label{table}
\setlength{\tabcolsep}{0.1pt}
\begin{tabular}{c p{0.15\linewidth} p{0.15\linewidth} p{0.15\linewidth} p{0.15\linewidth} p{0.15\linewidth} p{0.15\linewidth}}
\toprule
\multicolumn{1}{c}{} & \multicolumn{3}{c}{\textbf{Without aggregation}} & \multicolumn{3}{c}{\textbf{With aggregation}} \\
\cmidrule(rl){2-4} \cmidrule(rl){5-7}
\textbf{} & \centering \textbf{Dice} & \centering \textbf{Precision} & \centering \textbf{Recall} & \centering \textbf{Dice} & \centering \textbf{Precision} & \textbf{Recall}\\
\midrule
\centering $\begin{array}{c}  \textbf {Axial} \\
\end{array}$ & $0.7206 \pm 0.1369$ & $0.7133 \pm 0.1362$ & $0.7133 \pm 0.1364$ & $0.8272 \pm 0.0883$ & $0.7522 \pm 0.1889$ & $0.6584 \pm 0.1745$ \\
 \centering $\begin{array}{c}  \textbf {Sagittal} \\
\end{array}$ & $0.8051 \pm 0.0977$ & $0.6071 \pm 0.0884$ & $0.5853 \pm 0.0918$ & $0.8910 \pm 0.0685$ & $0.6509 \pm 0.1236$ & $0.5345 \pm 0.1306$ \\

 \centering $\begin{array}{c}  \textbf {Coronal} \\
\end{array}$ & $0.7308 \pm 0.1517$ & $0.7477 \pm 0.1659$ & $0.7301 \pm 0.1705$ & $0.8654 \pm 0.0859$ & $0.7664 \pm 0.1984$ & $0.6704 \pm 0.2191$ \\

\bottomrule
\end{tabular}
\label{tab3}
\end{table*}

\begin{table}[t]
\centering
\caption{The Report on the Combination of the Proposed Classification and Segmentation Models and Their Comparative Analysis with Other State-of-the-Art Models in Ischemic Lesion Segmentation, Including Models Trained on Both Lesion and Normal Slices}
\label{table}
\setlength{\tabcolsep}{0.3pt}

\begin{tabular}{p{0.15\linewidth} p{0.22\linewidth} p{0.1\linewidth} p{0.1\linewidth} p{0.1\linewidth} p{0.1\linewidth} p{0.12\linewidth} p{0.1\linewidth}}

\hline \\ \textbf{Modality} & \textbf{Methods} & \centering \textbf{Pre} & \centering \textbf{Rec} & \textbf{F1-score} & \textbf{Acc} & \textbf{  Sen} & \textbf{Spec} \\
\hline \\ \multirow{7}{*}{\centering \raisebox{-0.05\height}{\rotatebox{90}{\textbf{AXIAL}}}} & \textbf{VGG} & 95.35 & 95.35 & 95.35 & 95.35 & 95.404 & 95.28  \\ \\

& \textbf{Classification}  & 92.88 & 92.88 & 92.88 & 92.88 & 92.68 & 93.06  \\ \\

& \textbf{Combination} &  95.67 & 95.52 & 95.52 & 95.52 & 92.68 & 98.35  \\ \\


\hline  \\  \multirow{6}{*}{\centering \raisebox{-0.01\height}{\rotatebox{90}{\textbf{ SAGITTAL}}}} & \textbf{VGG} & 93.32 & 93.29 & 93.29 & 93.29 & 94.63 & 91.95 \\ \\

& \textbf{Classification}  & 91.25 & 91.24 & 91.24 & 91.24 & 91.86 & 90.61 \\ \\

& \textbf{Combination} &  95.12 & 94.95 & 94.94 & 94.95 & 91.86 & 98.02 \\ \\

\hline  \\  \multirow{6}{*}{\centering \raisebox{-0.01\height}{\rotatebox{90}{\textbf{ CORONAL}}}} & \textbf{VGG} & 95.31 & 95.31 & 95.31 & 95.31 & 96.59 & 96.03 \\ \\

& \textbf{Classification}  & 92.01 & 91.98 & 91.98 & 91.98 & 93.35 & 90.60 \\ \\

& \textbf{Combination} &  95.75 & 95.66 & 95.65 & 95.66 & 93.35 & 97.95 \\ \\

\hline
\end{tabular}

\label{tab3}
\end{table}

\subsection{classification and segmentation}
As described in the proposed method section, the final model is a combination of classification, segmentation, and aggregation function.

The performance of the model is assessed on a subset of 544 cases, comprising both lesion and normal slices that were not seen during training. The evaluation results of this approach are provided in Table 7. The model's performance is then compared to another state-of-the-art approach, which has been trained on both lesion and normal slices. Notably, the other state-of-the-art approach in the segmentation comparison has only been trained on lesion data and slices.

\subsection{Classification and segmentation with/without aggregation function}

The proposed method involves dividing 3D MRI images into 2D images for training and testing. In the final stage, an aggregation function is used to make segmentation decisions based on predictions from three separate planes. Each plane is trained independently to predict stroke lesions. The purpose of the aggregation function is to address the potential lack of clarity in individual planes when identifying lesions. Sometimes, the segmentation model may incorrectly classify certain brain regions as lesions, leading to false boundaries. By considering results from multiple planes, the aggregation function determines if a pixel predicted as a lesion in one plane is also identified as a lesion in the other planes. When the other planes agree on the presence of a lesion, the aggregation function confirms that pixel as a lesion. This approach allows the model to consider MRI images, the brain, and lesions from different perspectives, improving its overall understanding and accuracy. In this comparison, the 131 MRI 3D images, as described in the dataset section, are analyzed, with the results detailed in Table 12.


\subsection{act as a classification}
In the proposed methodology, the segmentation model is employed to delineate the regions in the brain slices where the classification model predicts the presence of a lesion. However, in cases where the classification model erroneously predicts the existence of a lesion in slices that do not actually contain any lesion, the segmentation model generates a ground truth that is entirely zero. This implies that the segmentation model fails to segment any part of the brain as a lesion, despite being trained on 2D slices that do contain one or more lesions. Consequently, when considering the combination of classification and segmentation as a unified classification model, the segmentation component serves to rectify false positive predictions made by the classification component, thereby enhancing the overall performance of the classification system. The outcomes of the model's classification performance are delineated in Table 13. Table 13 conveys that the classification model, in isolation, exhibits a comparatively inferior performance when contrasted with VGG. Nevertheless, when employed in conjunction with the segmentation model, the classification model outperforms the VGG model in terms of performance.

\section{Concolusion}
In conclusion, the recognition and accurate delineation of ischemic lesions in 3D magnetic resonance imaging scans represent a critical challenge in healthcare, with significant implications for diagnosis and treatment planning. This study introduces a novel VRU-net architecture, combining visual, Residual connection, and U shape networks, to address this challenge. The methodology incorporates a modified VGG model for lesion identification in individual 2D slices, followed by a U shape model with residual blocks for lesion segmentation within each 2D slice. This process is independently applied to axial, sagittal, and coronal planes, with the final segmentation achieved through aggregation. The segmentation model, on occasion, may erroneously classify certain brain regions as lesions, leading to the creation of false boundaries. To rectify this, the aggregation function diligently analyzes the predictions from multiple planes. It identifies instances where a pixel predicted as a lesion in one plane is also recognized as a lesion in the other planes. Only when there is a consensus among the various planes regarding the presence of a lesion does the aggregation function confirm the pixel as a lesion.
The adoption of a sequential approach, complemented by a classification model, enhances the Segmentation model's performance, while the decomposition of 3D images into 2D slices both simplifies the model's complexity and ensures a more precise segmentation by considering all three planes. Training on the Anatomical Tracings of Lesions After Stroke dataset demonstrates that the proposed model outperforms other state-of-the-art models in terms of accuracy and dice coefficient. Additionally, the Segmentation model provides valuable feedback to the classification model, contributing to a reduction in false positive predictions.

\end{document}